# Extreme Metasurfaces Enable Targeted and Protected Wireless Energy Transfer

Esmaeel Zanganeh[1], Polina Kapitanova[1], Andrey Sayanskiy[1], Sergey Kosulnikov[2], and Alex Krasnok[3,*]

[1]*School of Physics and Engineering, ITMO University, Saint-Petersburg, 197101, Russia*

[2]*Department of Electronics and Nanoengineering, Aalto University, FI-00079 Aalto, Finland*

[3]*Department of Electrical and Computer Engineering, Florida International University, Miami, FL 33174, USA*

[*]*e-mail:* [akrasnok@fiu.edu](akrasnok@fiu.edu)

**Abstract**

*Controlling the electromagnetic properties of materials beyond those achievable with natural substances has become a reality with the advent of metamaterials, artificially designed materials offering a wide range of unusual physical phenomena*[1–8]. *The extreme properties that metamaterials provide can protect optical and electromagnetic systems from surrounding "ordinary" materials and substances – a feature never explicitly used yet. Wireless energy transfer, i.e., the transmission of electromagnetic energy without physical connectors*[9], *demands reliable and stable solutions for charging high-power systems like electric vehicles with no effect on people, animals, plants, etc. Here we tackle this challenging problem and suggest a novel approach of using metamaterials with extreme parameters to enable targeted and protected wireless energy transfer. We design and experimentally implement epsilon-near-zero (ENZ) and epsilon-and-mu-near-zero (EMNZ) metamaterials*[10] *that provide an energy transmission if and only if both the transmitter and the receiver are equipped with these metamaterials. The fact of absence of materials with such extreme parameters protects the system against surrounding objects, which cause neither noticeable change in the system operation nor experience any detrimental effect. The system behind the proposed approach can be realised in virtually any frequency band by appropriate scaling and suitable choice of material. This technology will find applications in targeted wireless energy transfer systems, especially where high power is needed, including electric vehicles*[11].

**Main**

Metamaterials are artificially designed materials enabling material properties unachievable or barely found in nature. Many wisely designed metamaterials, consisting of identical or gradually changing cell arrays called artificial atoms or meta-atoms, were put forward, offering a wide range of unusual phenomena[1–8,12]. In particular, in electromagnetics and optics, metamaterials have enabled negative refractive index[1,2,13], superlenses[14,15], cloaking[3,16] nontrivial topological phases[17–19], to name just a few. Although the use of 3D metamaterials is limited by their losses and complexity of manufacturing, especially in optics, their 2D counterparts – metasurfaces – appear to be more loss-tolerant and feasible, often preserving the functionality of metamaterials, including control over the light propagation, reflection, and refraction[20–34].

Since the unusual parameters of metamaterials virtually never occur in nature, this circumstance could be used for control or protection – a property that has not been explicitly used yet. One can think of a system that performs desired functionality in the "metamaterial mode" but does not work otherwise. For example, in wireless power transfer (WPT), technology that enables charging a battery of portable and mobile devices without any additional plug-ins[9,35], a challenging problem is yet to be resolved – targeted and safe transfer of strong AC field energy without affecting surrounding objects. Indeed, the harmfulness of strong high-frequency electric fields explains the almost exclusive use of magnetic coupling[36] in lieu of electrical capacitive coupling[37]. On the other hand, although the magnetic WPT systems are designed to provide a move-and-charge scenario[35,38], they are limited to low-power systems due to harmful eddy currents heating metallic elements of, e.g., an electric vehicle[39]. Despite existing efficient low-power WPT solutions[36,40,41], the technology still demands systems robust to dynamic operating conditions[42] and capable of targeted energy transfer from a transmitter to randomly oriented and arranged receivers. If the WPT system worked only in the presence of a metamaterial with extreme parameters, then all surrounding objects would be protected by the absence of materials with such properties in nature, Fig. 1a.

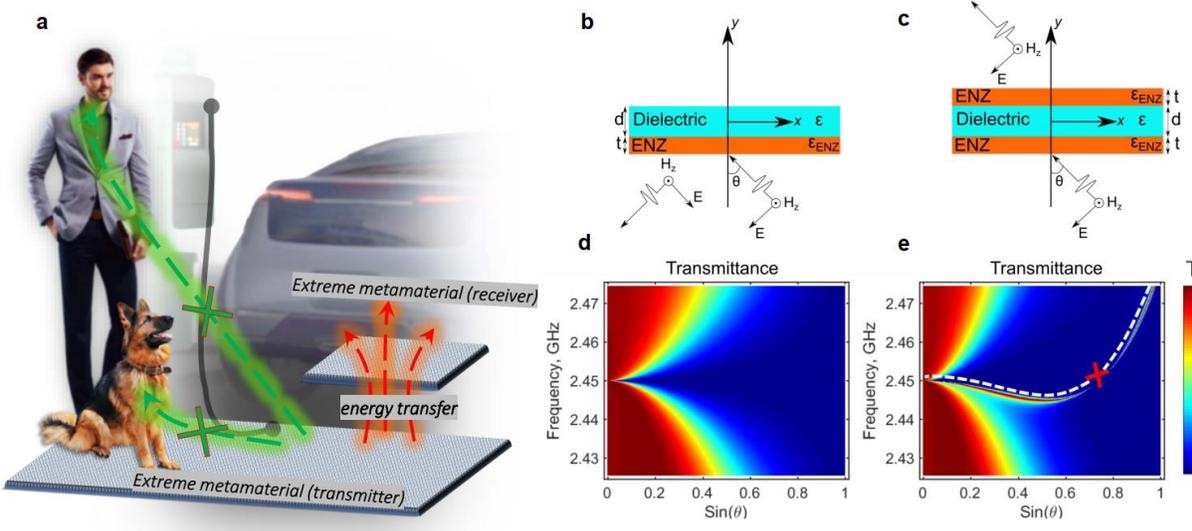

**Fig. 1| Illustration of the concept of extreme material-enabled targeted and protected wireless energy transfer. a,** Concept of wireless energy transfer targeted and protected by metamaterials with extreme parameters: the energy transmission occurs only between two extreme metamaterials, and the nonexistence of these extreme materials in nature protects the system from the presence of "ordinary" objects. **b, c,** Schematic view of transversely homogenous ENZ-dielectric (**b**) and ENZ-dielectric-ENZ (**c**) structures under TM-polarized plane wave illumination. **d, e,** Calculated transmittance of the ENZ-dielectric (**d**) and ENZ-dielectric-ENZ (**e**) structures versus frequency and angle of incidence $\theta$. The permittivity of the dielectric is $\varepsilon = 5$, its thickness is $d = 29$ mm, the thickness of the ENZ layers is $t = 2.9$ mm, and the plasma frequency is $f_p = 2.45$ GHz. The red cross in (**e**) denotes the BIC state.

In this paper, we propose a novel approach of targeted WPT involving metasurfaces with extreme parameters. The approach relies on two extreme-parameter metasurfaces and works as a WPT system when the two metasurfaces are stacked together. We develop an epsilon-near-zero (ENZ)[10] based multilayered structure illustrated in Fig. 1. In this scenario, the energy transmission occurs only between two extreme metamaterials, and the nonexistence of these extreme materials in nature protects the system from the presence of surrounding objects. We design and experimentally realise zero-index metamaterials with this functionality and demonstrate that the field does not penetrate through areas not covered by the upper metasurface layer providing targeted energy transfer.

It is worth noting that other "active" methods of targeted WPT have been proposed, including time division multiple access[42], multifrequency[43], and phase-shifted control[44]. These methods rely on complicated active control circuits with enhanced power consumption. Another active approach of on-site wireless power generation[45], while working only in the presence of the receiver circuit and consuming low energy in the passive mode, is very limited in its scalability and size. Moreover, several receivers would complicate these systems, making the existing approaches very challenging.

We begin with a rigorous analytical study of the electromagnetic wave tunnelling effect in two systems with extreme parameters: two-layered ENZ-dielectric heterostructure and three-layered ENZ-dielectric-ENZ heterostructure illustrated in Figs. 1b,c, respectively. This analysis will guide our design strategy to realise targeted WPT based on zero-index metamaterials. Both structures are illuminated by a TM-polarized plane wave with the magnetic field parallel to the structure interfaces. Such a structure can be realised in virtually any frequency band by appropriate scaling and suitable choice of material. In this paper, we consider the GHz spectral range without loss of generality. The permittivity of the ENZ material follows the Drude dispersion $\varepsilon_{ENZ}/\varepsilon_0 = 1 - \omega_p^2/(\omega^2 + i\gamma\omega)$, where $\omega = 2\pi f$ angular frequency, $\varepsilon_0$ is the free-space permittivity, and $\omega_p$ and $\gamma$ are, the plasma frequency and collision frequency of the electron gas, respectively. We first consider the lossless case $\gamma = 0$ and set the plasma frequency to $f_p = \omega_p/2\pi = 2.45$ GHz. We calculate the transmittance through the systems for different incidence angles $\theta$ of the plane wave.

The ENZ-dielectric structure (from now on, 2L-structure) containing one ENZ layer (Fig. 1b) under TM-polarized excitation reveals angle-dependent and dispersive transmission except for the plasma frequency, where it vanishes, Fig. 1d. The vanishing of transmission at the plasma frequency is explained by diverging transverse magnetic (TM) wave impedance along the z-direction in ENZ $Z_{ENZ} = \eta_0 \cos\theta_1 / \sqrt{\varepsilon_{ENZ}}$ when $\varepsilon_{ENZ}$ tends to zero. Here $\eta_0 = \sqrt{\mu_0/\varepsilon_0}$, and $\theta_1$ is the angle of refraction in the ENZ layer obtained using Snell's law. Hence, the field exciting this ENZ layer reflects back to the transmitter antenna. Notable, as the incident angle approaches zero, the Brewster condition corresponding to the unit transmission gets to the plasma frequency. It merges with the zero transmission point enabling the bound state in the continuum (BIC) at normal incidence[46–48]. BIC manifests itself as a visible sharp resonant feature at the plasma frequency, Fig. 1d.

Adding a second ENZ layer (from now on, 3L-structure), as shown in Fig. 1c, enables the field tunnelling mode that manifests itself as a narrow perfect transmission resonance in the spectrum around the plasma frequency 2.45 GHz, Fig. 1e. In the tunnelling mode, the 3L-structure allows the electromagnetic wave of a proper frequency and angle of incidence to tunnel from the first ENZ layer through the dielectric to the second ENZ layer. This tunnelling depends on the effective permittivity of the 3L-structure and becomes possible only in the presence of both ENZ layers. This mode appears at the intersection of the Fabry-Perot mode of the dielectric layer and the ENZ resonances in the claddings. In ref.[46] this mode was found to be associated with the so-called Berreman mode mainly studied in metal-dielectric multilayers[49–51] and primarily explained through the effective medium theory[52,53]. This mode in spectrum makes the structure perfectly transparent, providing the wave tunnelling through the heterostructure for all incident angles. The same behaviour is obtained for an incident TE-polarised wave in mu-near-zero (MNZ) configuration (Supplementary Materials, Section I). The most attractive case for applications capable of operating with both TE and TM polarisations would be the epsilon-and-mu-near-zero metasurface (EMNZ) that we also investigate (Supplementary Materials, Section II) and realise experimentally.

We also observe the accidental BIC state[54] appearing in the spectrum as an unboundedly narrowing transmission and reflection line marked by the red cross in Fig. 1e. A true BIC with infinite radiative Q-factor and vanishing resonance width can exist only in ideal lossless infinite structures or for extreme values of parameters[55–57], as in the considering case. In the 3L-structure, this anomaly scattering state appears when the Fabri-Perot mode of the dielectric spacer layer coincides with the ENZ resonances of the ENZ layers; hence the dielectric mode becomes trapped[58]. In what follows, we design and realise the extreme-parameter metasurfaces and demonstrate theoretically and experimentally the tunnelling effect for WPT.

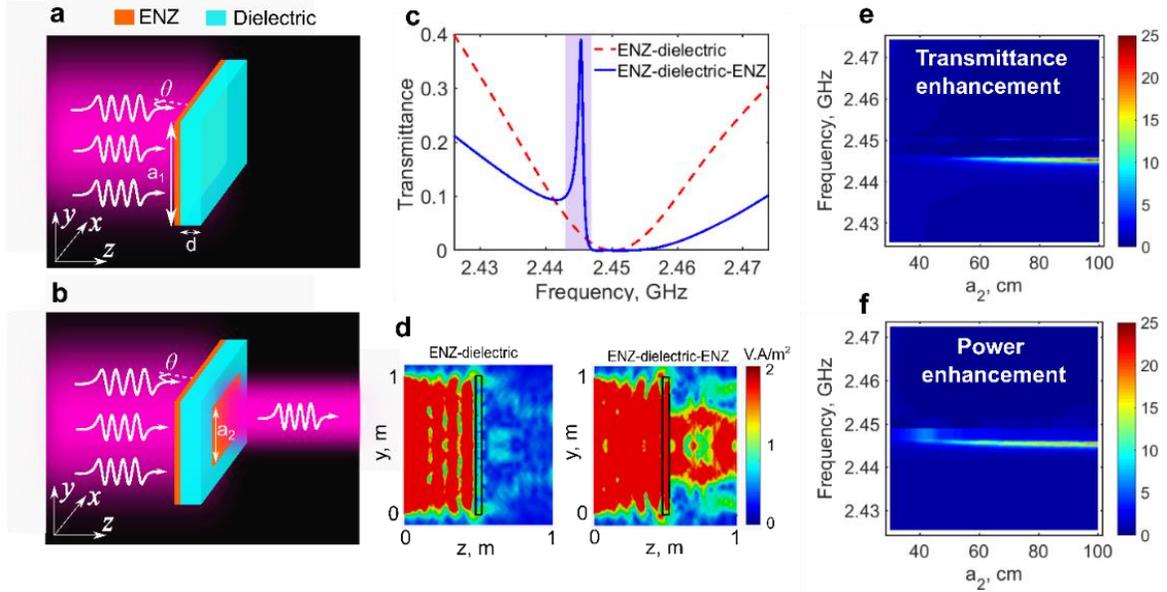

**Fig. 2| Tunneling in a finite structure. a, b,** Schematic view of the finite (**a**) ENZ-dielectric and (**b**) ENZ-dielectric-ENZ structures under TM-polarized plane wave illuminations. The second ENZ (plate) is smaller than the first one (layer). The size of the dielectric and ENZ layer is $a_1 = 100$ cm. The second square ENZ size is $a_2 = 70$ cm. Other parameters are the same as in Fig. 1. The incident wave angle is set to 30°. **c,** Simulated transmittance of the 2L- and 3L-structures. **d,** Simulated power flow through the structures in panels (**a**) and (**b**) at plasma frequency. **e,** Transmittance of the 3L-structure normalised to the one of 2L-structure versus frequency and $a_2$. **d,** Integrated power flow over the surface in x-y plane 4 cm above the ENZ plate and normalised to the same without a plate.

Now we investigate the tunnelling effect in the finite structures, Fig. 2. The first ENZ layer (from now on, a layer) and dielectric are now square sheets of size $a_1 = 100$ cm. The second ENZ layer (from now on, a plate) is a square sheet of size $a_2 = 70$ cm. In Fig. 2c, we compare the transmittance of 3L- and 2L-structures versus frequency for the incident wave angle of $\theta = 30^o$. For 3L-structure we observe the transmittance of 0.4 and a minor frequency deviation with respect to the infinite case due to the finite size of the structure. We normalised transmittance $T_{3L}/T_{2L} \simeq 8$ at the plasma frequency. Remarkably, in the 3L-structure, the energy of the electromagnetic field indeed propagates majorly through the plate. Otherwise, the field does not pass the structure, as revealed by simulations, Fig. 2d. To elucidate the effect of size, Fig. 2e shows the transmittance versus frequency normalised to the case of no ENZ plate. The enhancement gradually increases for $a_2 > 50$ cm and reaches 25 for $a_2 = 100$ cm. Fig. 2f shows the power flow integrated over the surface at 4cm over the ENZ plate and normalised to the case of no plate. The maximum integral power enhancement of 25 is observed, proving that the energy mostly goes through the ENZ plate.

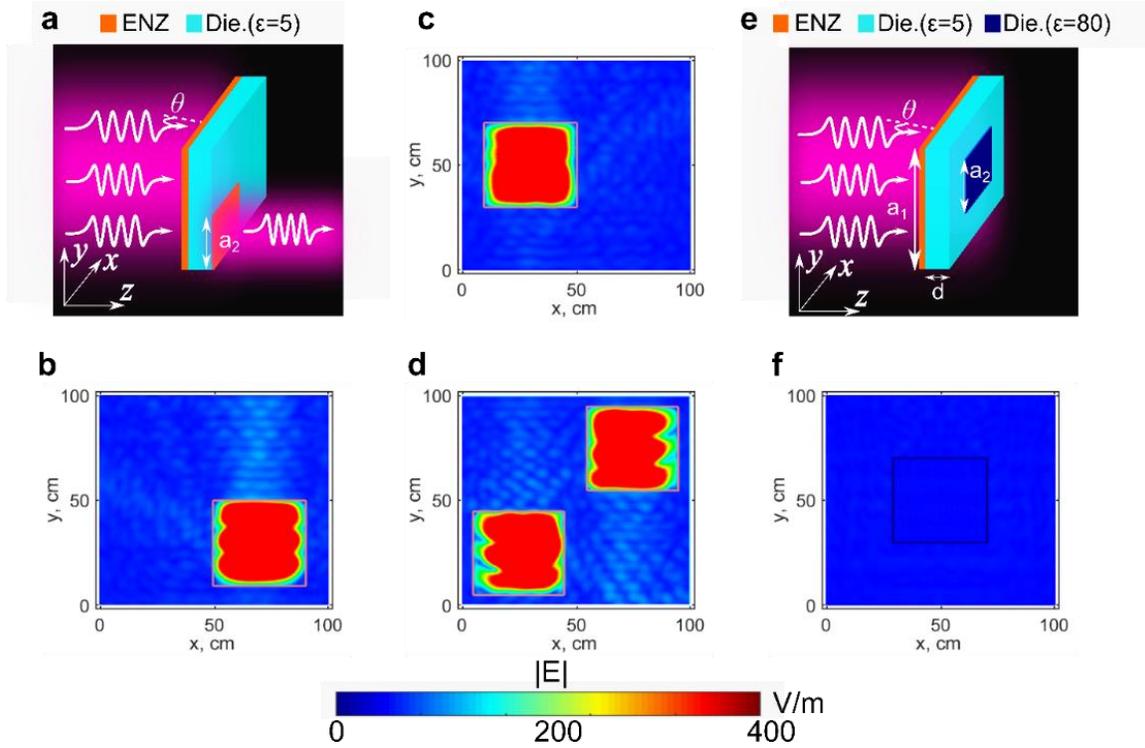

**Fig. 3| Targeted power transfer in the finite structure. a,** Finite 3L-structure with shifted ENZ plate. **b, c,** Calculated electric field distribution of the finite 3L-structure for different locations of the ENZ plate. **d,** Calculated electric field distribution for the presence of two ENZ plates. **e,** Finite 3L-structure with ENZ plate replaced by a dielectric with permittivity $\varepsilon = 80$ of the same size. **f,** Calculated electric field distribution of the 3L-structure with a dielectric plate showing vanishing transmission.

An arbitrary displacement of the ENZ plate over the structure's surface reveals that the tunnelling effect follows it, Fig. 3a. For example, Figs. 3b and c illustrate this conclusion for two very different plate positions. The tunnelling effect, being enabled by the plate with extreme parameters, finds its way through this layer whenever it is located. Thus, by changing the position of the ENZ plate, one can easily guide the power tunnelling route. The strong enhancement of the electric field compared to the incident field (~400 times) over the area of the ENZ plate proves that the majority of tunnelled power concentrates within the ENZ plate. Moreover, two ENZ plates enable the power tunnel through both plates, Fig.3d. In order to demonstrate that for this effect to occur, the plate must be made of ENZ material, a substance that virtually never meets in nature, we simulate a plate of high permittivity dielectric ($\varepsilon = 80$) of the same size being placed instead of ENZ plate, Figs. 3e,f. Such dielectric slab mimics water and water-containing substances like living tissues in the radio and microwave frequency range. As we can see, the dielectric plate demonstrates no tunnelling effect.

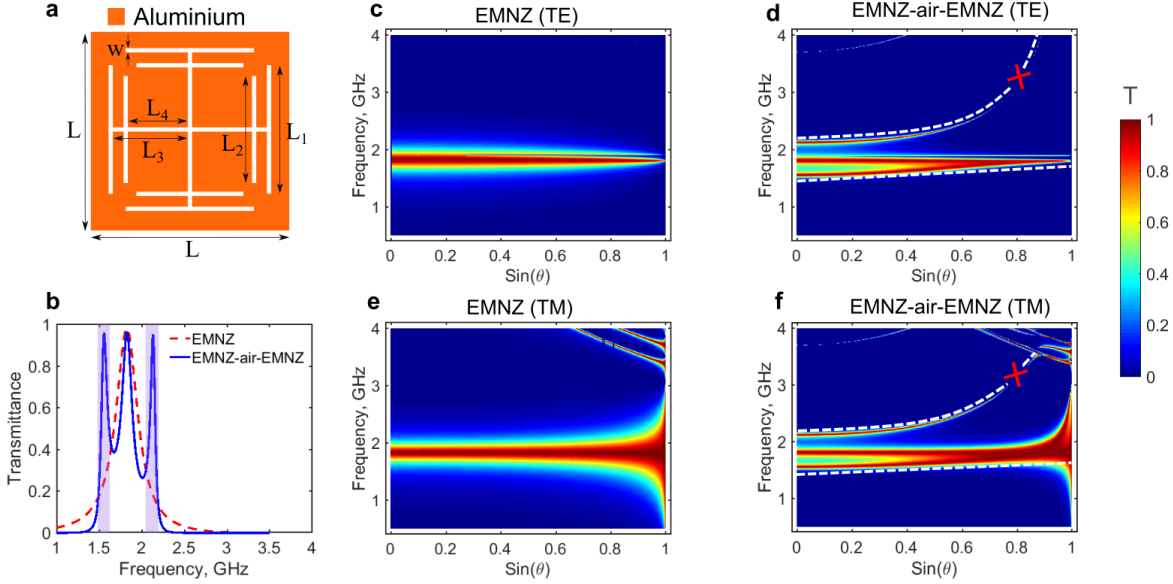

**Fig. 4| Tunneling in realistic metamaterial structure. a,** Geometrical parameters of the EMNZ metasurface unit cell: L=45 mm, $L_1$=28 mm, $L_2$=20 mm, $L_3$=16 mm, $L_4$=11.5 mm, w=1 mm and thickness=2 mm. **b,** Calculated transmittance (normal incidence) of the infinite EMNZ-air and EMNZ-air-EMNZ structures made of this metasurface. The air gap in the EMNZ-air-EMNZ structure is d=80 mm. **c, d,** Transmittance of the (**c**) EMNZ-air and (**d**) EMNZ-air-EMNZ infinite structures as a function of the frequency and incident wave angle for TE-polarised waves. **e, f,** Similar to (**c**) and (**d**) but for TM-polarized illumination.

We implement the structures introduced above using a realistic aluminium metasurface design illustrated in Fig. 4a. The metasurface has a quadratic unit cell to take advantage of both polarisations. We extract the effective permittivity and permeability using the well-known retrieval method[59,60] (Supplementary Materials, Section II). The results of extraction are presented in Fig. S2c,d. The designed metasurface possesses an ENZ response in the frequency range 1–3 GHz, with the plasma frequency at ~1.83 GHz. This metasurface's design provides virtually constant permeability of μ=0.07+i0.002 in this range. Both effective parameters exhibit a Lorentz-type resonance at the frequency f=3.3 GHz. Thus, the proposed metasurface acts as an EMNZ media over the 1–3 GHz frequency band. To reduce the loss factor in the dielectric, we use either air (in simulations) or microwave foam (in experiments) with $\varepsilon_d \simeq 1$ and vanishing loss.

To validate the tunnelling effect of the proposed EMNZ-air-EMNZ structure, we calculate the wave transmittance through the 2L- and 3L-strictures (Supplementary Materials, Section II). Fig. 4b shows the transmittance through a single EMNZ layer (red dashed curve). The transmission reaches 0.95 at f=1.83 GHz, where the effective wave impedance $Z_{EMNZ} = \sqrt{\mu_{EMNZ}/\varepsilon_{EMNZ}}$ is almost matched with air. In this realistic system, the maximal transmittance is reduced due to the material losses of the metal. The presence of the second EMNZ leads to $T_{3L}/T_{2L} \simeq 4$ at 1.56 GHz and 2.13 GHz, Fig. 4b (solid blue curve). The transmittance of the 2L- and 3L-structures for the TE- and TM-polarized waves as a function of incident angle and frequency are shown in Fig. 4c-f. The comparison of panels (d) and (f) with panels (c) and (e) reveals that the presence of the second EMNZ layer leads to two narrow transmission lines (indicated by dashed white lines). The formation of these transmission modes is attributed to the strong coupling effect and analysed by coupled modes approach (see Methods). Remarkably the appearance of these modes makes the structure transparent for almost all incident angles except the angle associated with BICs (red cross).

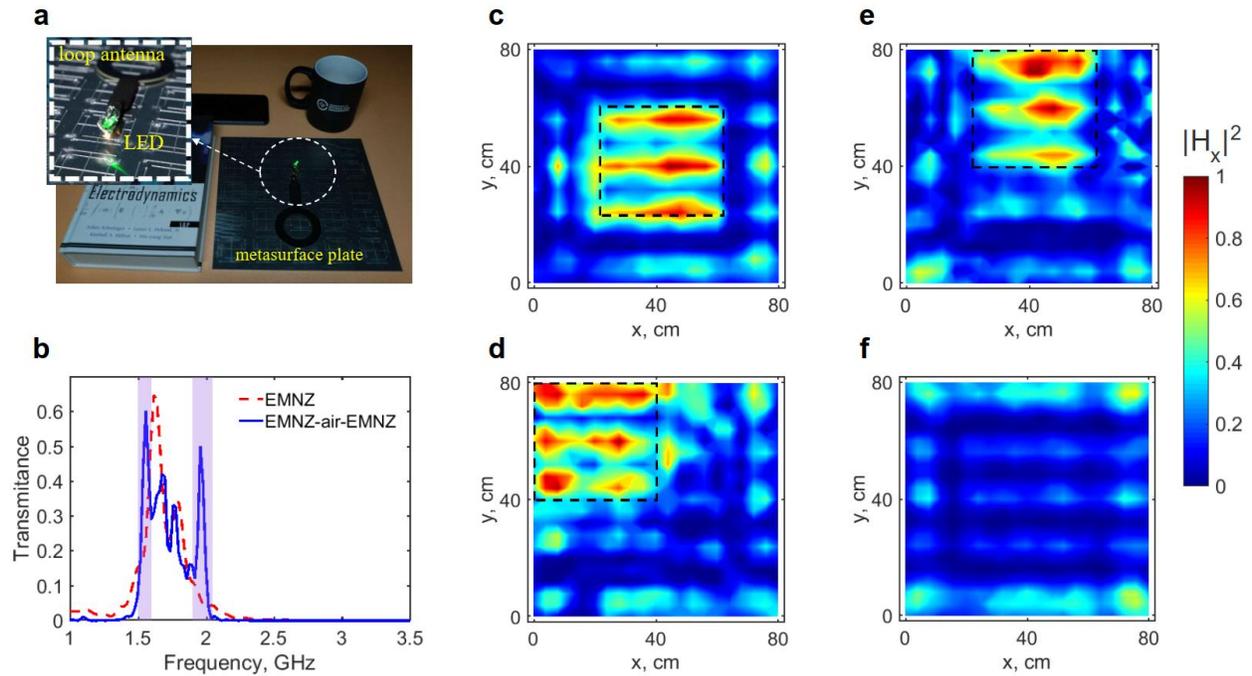

**Fig. 5| Experimental demonstration of the targeted power transfer. a,** LED demonstration of the targeted power transfer in the 3L-structure. The lower EMNZ metasurface is 81×81 cm² (18×18 cells) in all experiments. The distance between two EMNZ metasurfaces is d=80 mm. The LED demonstration experiment is conducted for the EMNZ plate of 22.5×22.5 cm² (5×5 cells) size. **b,** Measured transmittance of the 2L- and 3L-structures (two 18×18 cells metasurfaces). **c-e,** Measured magnetic field of the 3L-structure in the x-y plane at 10 cm over the EMNZ plate of 40.5×40.5 cm² (9×9 cells) for different positions: (**c**) at the centre, (**d**) at the edge, and (**e**) at the corner. **f,** Measured magnetic field of the EMNZ-air structure in the x-y plane at 10 cm over. Fields are normalised to the maximum in panel (**c**) in all cases**.**

The prototypes of 2L and 3L -structures using EMNZ metasurfaces were fabricated and experimentally investigated. We measured the transmittance through the 2L- and 3L-structures and the near magnetic field distributions over the EMNZ plate (see Methods and Supplementary Materials for details). We directly visualised the power flow using a receiver loaded with a light-emitting diode (LED). The structures under study were excited by a horn antenna to mimic a plane wave of different polarisation. We first demonstrate the targeted WPT in the 3L-structure using a wideband loop receiver above the EMNZ plate. The current induced in the loop antenna was rectified and delivered to the LED, Fig. 5a. The LED is ON while the receiver is located above the EMNZ plate, even in the presence of other objects, no matter how complex they are. It proves the tunnelling of electromagnetic waves and targeted power transfer through the 3L-structure. When the LED is placed in another place but plate, it is OFF, indicating no energy transfer. To further elucidate the transmission enhancement, we measure transmittances of the EMNZ-air and EMNZ-air-EMNZ structures and compare them in Fig. 5b (we refer to Fig. S3c in Section II of Supplementary Materials for simulation results). We observe the transmission enhancement as much as $T_{3L}/T_{2L} \simeq 3$ and $T_{3L}/T_{2L} \simeq 10$ at 1.56 GHz and 1.96 GHz, respectively. Although the results are in excellent agreement with the numerical results for both infinite (Fig. 4) and finite structures (Fig. S3), we also observe the appearance of several minor modes caused by the unit cells eigenmodes and their strong interlayer interactions.

Finally, we measure the near magnetic field distribution in the x-y plane 10 cm over the 3L-structure for different locations of the EMNZ plates (see Section III of Supplementary Materials). As illustrated in Fig. 5c-e, the magnetic field is strongly concentrated in the area of the EMNZ plate. The measured field distribution of the 2L structure presented in Fig. 5f reveals a vanishing magnetic field everywhere in the absence of the plate.

We have demonstrated a novel approach of targeted WPT involving metasurfaces with extreme parameters. The approach uses two extreme-parameter metasurfaces and allows energy transmission if and only if two metamaterials are stacked together. We have experimentally realised this approach with EMNZ-air-EMNZ metastructures and demonstrate that the field does not penetrate through areas not covered by the smaller extreme metasurface plate. This suggests a targeted energy transfer enabled and protected by the extreme parameter metamaterials – wherever we place the metamaterial plate, it enables energy tunnelling and concentration. The presence of other objects with "ordinary" parameters, regardless of their complexity, including electronic devices, people, animals, plants, etc., leads to neither any noticeable change in the functionality of the WPT system nor any detrimental effect. We believe that this technology can be applied in targeted wireless energy transfer systems. Of particular importance are systems where high power is needed, including charging electric vehicles[11]. This implies that the parking spot is covered with one layer of the extreme metasurface and the second layer (plate) is set on the vehicle. The transmitting antenna feeding the first layer and the receiving antenna nearby the metasurface plate can be conventional antennas, making this technology more feasible.

**Supplementary information**

This file contains the results on tunnelling in an infinite MNZ-dielectric-MNZ structure (TE-polarization), design of metasurface-based EMNZ structures, details on the effective parameters extraction, details on prototype fabrication and experimental study of the targeted power transfer and Figures S1-S5.

**Methods**

**Transfer matrix formalism**

For given parameters of extreme materials and dielectric $\varepsilon_{ENZ}$, $\mu_{MNZ}$, and $\varepsilon$, we define the transfer (T-) matrix for the complete layered structure[61], $\hat{\mathbf{T}}_t = \hat{\mathbf{T}}_m \otimes \hat{\mathbf{T}}_d \otimes \hat{\mathbf{T}}_m$. Here $\hat{\mathbf{T}}_m$ and $\hat{\mathbf{T}}_d$ are T-matrices for the extreme materials and dielectric layer, respectively, $\hat{\mathbf{T}} = \hat{\mathbf{D}} \otimes \hat{\mathbf{P}} \otimes \hat{\mathbf{D}}$, where matrices $\hat{\mathbf{D}}$ describe propagation across the boundary between layers i and j, and the matrix $\hat{\mathbf{P}}$ accounts for propagation through the particular layer. For the TE-polarisation (s) and TM-polarisation (p), these matrices can be expressed in the form: $\hat{\mathbf{D}}_{ij}^{s,p} = \frac{1}{2}\begin{bmatrix} 1+\eta_{s,p} & 1-\eta_{s,p} \\ 1-\eta_{s,p} & 1+\eta_{s,p} \end{bmatrix}$, and $\hat{\mathbf{P}}_i = \begin{bmatrix} \exp(-ik_z^i d^i) & 0 \\ 0 & \exp(ik_z^i d^i) \end{bmatrix}$, where $\eta_s = k_z^{i+1}/k_z^i$, $\eta_p = \varepsilon_i k_z^{i+1}/\varepsilon_{i+1} k_z^i$, $k_z^i$ is the out-plane wave number in the particular layer i, $d^i$ is the thickness of the layer i. With the T-matrix of the whole structure $\hat{\mathbf{T}}_t \equiv \begin{bmatrix} T_{11} & T_{12} \\ T_{21} & T_{22} \end{bmatrix}$, we calculate the reflectance ($R$) and transmittance ($T$) spectra for the multilayer structure, $R = |T_{21}/T_{11}|^2$, $T = |1/T_{11}|^2$.

**Mode coupling description**

The appearance of two resonant transmission bands in the spectrum (Fig. 4d,f) is explained by the strong coupling of modes of the two EMNZ layers. The modes are characterised by the corresponding complex eigenfrequencies $\omega_{1,2} - i\omega_{1,2}/2Q_{1,2}$, where $\omega_{1,2}$ is the real frequency of the modes and $Q_{1,2} = \omega_{1,2}/2\Delta\omega_{1,2}$ is their quality factor defined by the spectral bandwidth $\Delta\omega_{1,2}$. The effective interaction Hamiltonian $\hat{H}_{int}$ of the two modes considered as two oscillators with the coupling strength $g$ yields:

$$\hat{H}_{int} = \begin{pmatrix} \omega_1 - i\omega_1/2Q_1 & g \\ g & \omega_2 - i\omega_2/2Q_2 \end{pmatrix}, \qquad (1)$$

The coupling strength depends on the mode overlap, and for the close spacing $d$ it is real-valued. The diagonalisation of this Hamiltonian yields the new eigenfrequencies of dressed modes, $\omega_\pm = (\omega_1 + \omega_2)/2 - i(\Delta\omega_1 + \Delta\omega_2)/2 \pm \sqrt{g^2 + [\delta - i(\Delta\omega_1 - \Delta\omega_2)]^2/4}$, where $\delta = \omega_1 - \omega_2$ is the detuning. In our case of identical metasurfaces, the mode splitting (Rabi splitting) is $\Omega = 2g$. The transmittance spectrum (solid blue curve) in Fig. 4b reveals the Rabi splitting $\Omega = 0.58$ GHz, corresponding to the coupling strength $g = 0.29$ GHz. The average mode widths of our system is $(\Delta\omega_1 + \Delta\omega_2)/2 = 0.1$ GHz. Thus, since g > $(\Delta\omega_1 + \Delta\omega_2)/2$, the modes are strongly coupled, giving rise to mode splitting.

**Numerical simulations**

The simulations in Fig. 1-4 and Supplementary Fig. S1-S3 were performed using CST Microwave Studio 2020. All simulations used extremely fine mesh-cell settings, determined by adaptive meshing results. All the parameters used in the CST simulations were close to experimentally realised systems. The simulations were performed using a frequency-domain solver.

The S-parameter-based method is used to retrieve the effective material parameters of the single-layer Metasurface (Section II of Supplementary Materials). Considering the normal incident wave, we first calculated effective refractive index $n$ and wave impedance Z of the metasurface from the S-parameters. Then permittivity ($\varepsilon$) and permeability ($\mu$) were calculated from calculated effective refractive index $n$ and wave impedance Z.

**Metamaterial fabrication**

Several EMNZ metasurface prototypes made of aluminium (conductivity $3.56\times10^7$ S/m) with the sizes of $81\times81$ cm$^2$ ($18\times18$ cells), $40.5\times40.5$ cm$^2$ ($9\times9$ cells), and $22.5\times22.5$ cm$^2$ ($5\times5$ cells) were fabricated using the laser cutting method. The slots width is $1\pm0.05$mm.

**Experimental methods**

To measure the transmittance of the wave irradiated by a horn antenna through the sample, the near magnetic fields in a scan area close to the surface of the prototype were measured and then, using transformation to the far-field zone and post-processing, the transmittances were obtained[62]. To measure the magnetic nearfield distribution for EMNZ-air and EMNZ-air-EMNZ structures with a small second EMNZ, the x component of the magnetic field ($H_x$) was scanned over the area of $80 \times 80$ cm$^2$ with 4 cm step and at the distance of 10 cm from the prototypes surface. More details can be found in SM section III.

**Acknowledgements**

The work is financially supported by the Program Priority 2030. EZ and PK acknowledge the technical support of Alexander Kalganov in the fabrication of the EMNZ metasurfaces and Mikhail Kuzmin for technical support in experiments.

**Author contributions**

A.K., E.Z., and P.K. conceived the idea. E.Z., S.K. and P.K. performed the simulations and advised in the experimental designs. E.Z. and P.K. led the experiments. E.Z., P.K., M.K., and A.S. performed the measurements. E.Z., A.K. and P.K. analysed the data. All authors contributed to the interpretation of the results. E.Z., A.K. and P.K. wrote the manuscript, with input and comments from all authors. A.K. supervised the project.

**Competing interests**

The authors declare no competing interests.

# Supplementary Materials: Extreme Metasurfaces Enable Targeted and Protected Wireless Energy Transfer

Esmaeel Zanganeh[1], Polina Kapitanova[1], Andrey Sayanskiy[1], Sergey Kosulnikov[2], and Alex Krasnok[3,*]

[1]School of Physics and Engineering, ITMO University, Saint-Petersburg, 197101, Russia

[2]Department of Electronics and Nanoengineering, Aalto University, FI-00079 Aalto, Finland

[3]Department of Electrical and Computer Engineering, Florida International University, Miami, FL 33174, USA

[*]e-mail: akrasnok@fiu.edu

**Section I. Tunneling through an infinite MNZ-dielectric-MNZ structure (TE polarization)**

As a complementary to discussion on the tunneling of electromagnetic waves with transverse magnetic (TM) polarization in ENZ-dielectric-ENZ structure, we investigate the tunneling of waves with transverse electric (TE) polarization (electric field parallel to the structure interfaces) in MNZ-dielectric-MNZ structure. Corresponding 2L- and 3L-structures are shown in Fig.S1 (a) and (b), respectively. Both structures are illuminated by a TE-polarized plane wave with the incident wave angle of θ. The permeability of the MNZ material follows a classic Drude dispersion $\mu_{\mathrm{MNZ}}/\mu_0 = 1 - \omega_p^2/(\omega^2 + i\gamma\omega)$, where $\mu_0$ is the free-space permeability, $\omega_p$ is the parameters plasma and $\gamma$ is the collision frequency. We consider the lossless case $\gamma=0$ and set plasma frequency as $\omega_p=2\pi f_p$, where $f_p$=2.45 GHz.

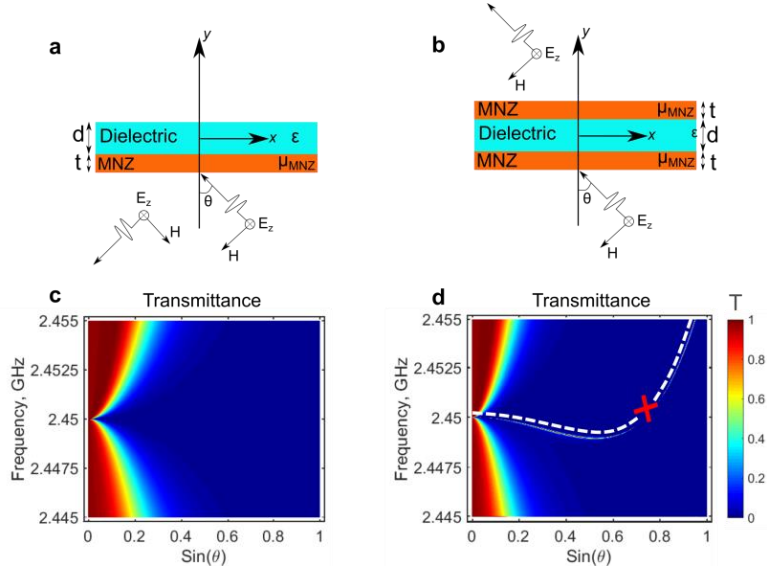

**Figure S1| Tunneling in an infinite MNZ-dielectric-MNZ structure.** Schematic view of (**a**) transversely homogenous *MNZ-dielectric* and (**b**) *MNZ-dielectric-MNZ* slabs under TE-polarized plane-wave illuminations. The

permittivity of the dielectric and its thickness are ε = 5 and d = 29 mm, respectively. The thickness of MNZ layers is t=2.9 mm. Calculated transmittance of the MNZ-dielectric structure **(c)** and MNZ-dielectric-MNZ structure **(d)**.

Transmittance of the 2L- and 3L-structures as a function of the incident angle and frequency are presented in Fig. S1 (c) and (d). Similar to the ENZ case, in the transmittance of MNZ-dielectric-MNZ structure, a sharp resonance with the unitary transmission is observed for all of the incident angles (indicated by the dashed white line) except the one corresponding to BICs (indicated by the red cross).

**Section II. Design of metasurface-based EMNZ structures**

Here, we discuss the simulation methods for calculating transmittance and effective permittivity and permeability of the epsilon-mu-near-zero (EMNZ) metasurface unit cell. The numerical model of the EMNZ-air and EMNZ-air-EMNZ unit cells constructed in CST Microwave Studio 2020 are depicted in Fig. S2 (a). Floquet ports and boundary conditions are used. The material of the unit cells is aluminum with the conductivity σ=3.56×10$^7$ S/m. The geometrical dimensions of the unit cell are as follows: L=45 mm, $L_1$=28 mm, $L_2$=20 mm, $L_3$=16 mm, $L_4$=11.5 mm, w=1 mm and thickness=2 mm. The numerical study of the unit cell models was performed with the frequency-domain solver of CST Microwave studio 2020 in the frequency range 1-3.5 GHz. During the simulation, the S-parameters of the models were obtained. The transmittance is defined as |$S_{21}$|$^2$, where $S_{21}$ is the transmission coefficient of the structure. The S-parameter-based method is used to retrieve the effective material parameters of the single-layer metasurface[1,2]. Considering the normal incident wave, we first calculated effective refractive index *n* and wave impedance Z of the metasurface from the S-parameters as follows[1]:

$$n = \frac{1}{kt}\cos^{-1}\left[\frac{1}{2S_{21}}\left(1 - S_{11}^2 + S_{21}^2\right)\right] \quad (1)$$

$$Z = \sqrt{\frac{(1+S_{11})^2 - S_{21}^2}{(1-S_{11})^2 - S_{21}^2}} \quad (2)$$

where, $S_{11}$ is the reflection coefficient, $S_{21}$ is the transmission coefficient, *k* is wave number, and *t* is the thickness of the metasurface. Then permittivity (ε) and permeability (μ) were calculated as:

$$\varepsilon = \frac{n}{Z} \quad and \quad \mu = nZ. \quad (3)$$

The obtained effective wave impedance, permittivity and permeability of the single-layer EMNZ metasurface are shown in Fig. S2 (c-e).

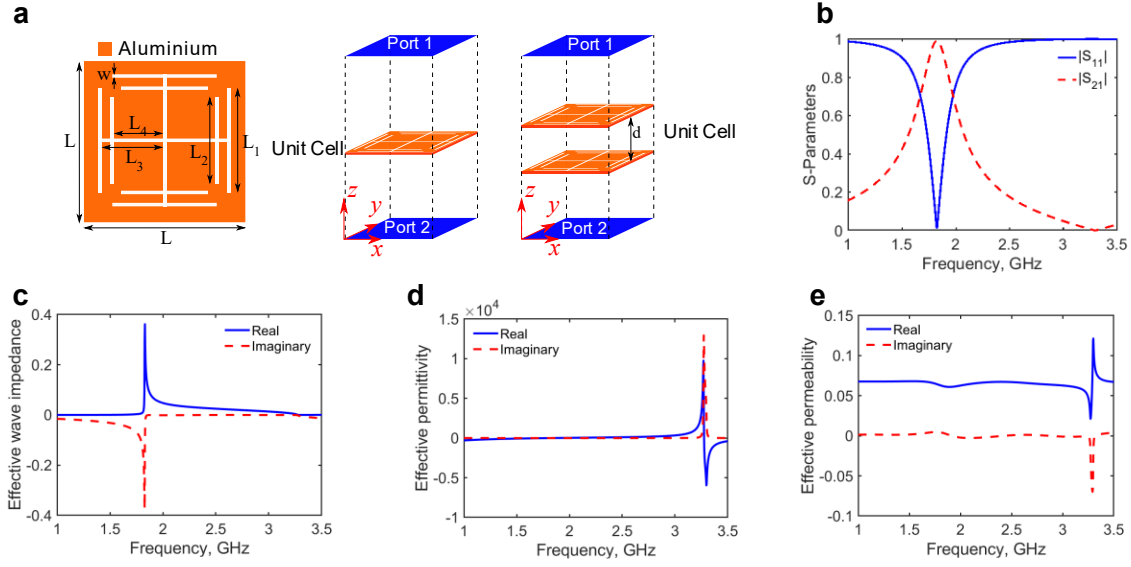

**Fig. S2| Metasurface-based EMNZ structures.** (**a**) Geometrical parameters and simulation setups for the single-layer and two layers EMNZ metasurface unit cells. The distance between two layers is d=80 mm. (**b**) Calculated S-parameters amplitude of the single-layer EMNZ metasurface unit cell. Real and imaginary part of the single-layer metasurface (c) effective wave impedance, (d) effective permittivity, and (e) effective permeability calculated from S-parameters using retrieval method.

The transmittance of the finite size EMNZ structures was also numerically simulated with the help of the frequency-domain solver of CST Microwave Studio 2020. The numerical models of EMNZ-air and EMNZ-air-EMNZ structures are shown in Fig. S3 (a)and (b), respectively. The metasurfaces are constructed from 18×18 unit cells and are characterized by the total size is 81×81 $cm^2$. The distance between metasurfaces in Fig. S3 (b) is d=80 mm. The waveguide ports are used to excite the structure and receive the transmitted power. The open boundary conditions are applied. The transmittance is defined as $|S_{21}|^2$, where $S_{21}$ is the transmission coefficient of the structures. The numerically obtained transmittances are compared in Fig. S3 (c). Four-fold and six-fold enhancements are observed in transmittance of the EMNZ-air-EMNZ structure at 1.56 GHz and 2.12 GHz, respectively, compared to the transmittance of the EMNZ structure.

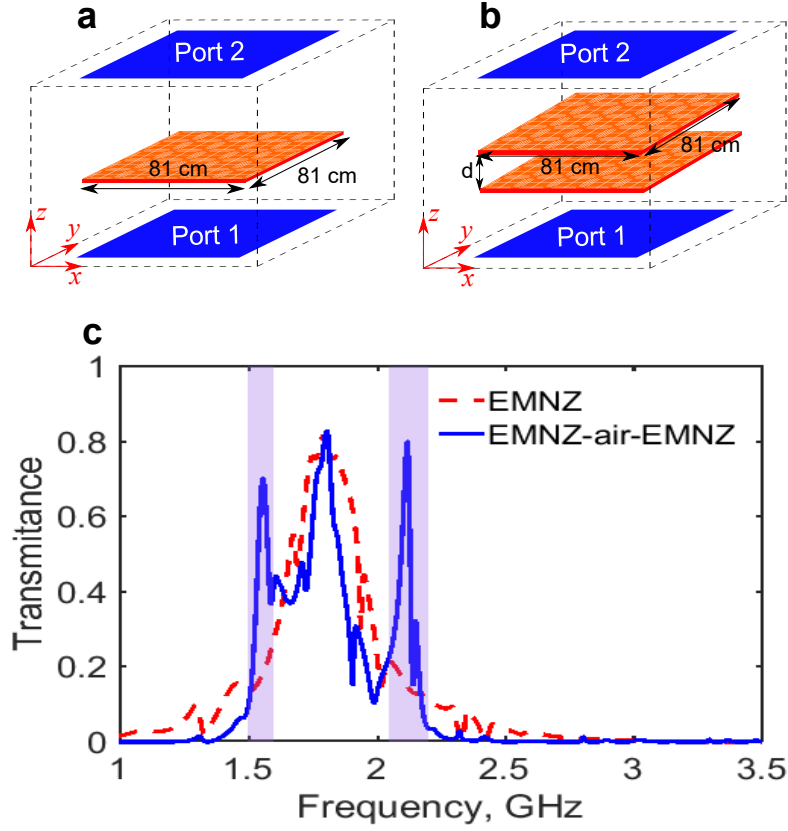

**Fig. S3| Transmittance of finite EMNZ-air and EMNZ-air-EMNZ structures.** Numerical models for (a) EMNZ - air structure *and (b) for EMNZ-air-EMNZ structure. Metasurfaces size is 81×81 cm$^2$. They are composed of 18×18 unit cells. (c) Calculated transmittances of EMNZ-air and EMNZ-air-EMNZ structures.*

**Section III. Prototype fabrication and experimental study of the targeted power transfer**

Several EMNZ metasurface prototypes made of aluminium (conductivity 3.56×10$^7$ S/m) with the sizes of 81×81 cm$^2$ (18×18 cells), 40.5×40.5 cm$^2$ (9×9 cells), and 22.5×22.5 cm$^2$ (5×5 cells) were fabricated using the laser cutting method. The geometrical dimensions of the unit cell are as follows: L=45±0.05mm, $L_1$=28±0.05 mm, $L_2$=20±0.05 mm, $L_3$=16±0.05 mm, $L_4$=11±0.05mm, w=1±0.05mm and thickness=2±0.025mm.

During the experimental study, we measured the transmission coefficient and near magnetic fields. We also performed a direct test of power tunnelling with the help of a receiver loaded by a light-emitting diode (LED).

To measure the transmittance of the wave irradiated by a horn antenna through the sample, we first tried to apply the far-field technique. But we found a problem in the post-processing of the measured data due to the high level of interference from the borders of the samples. It made this approach impossible to make transmittance measurements on a conventional far-field range. Thus, we applied an alternative approach to determine transmittance from measurements of the magnetic field made in the radiating near-field region[3], as shown in Fig. S4 (a) and (b). For that purpose, we placed the prototypes at 2.5 m distance from a horn antenna in the anechoic chamber. The horn antenna was connected to the first port of a Vector Network Analyzer (VNA) ZVB20 by a 50 Ohm coaxial cable. An electrically small magnetic probe mounted to an arm of the 3-axis scanner and connected to the second port of the VNA was used to detect the x component of the magnetic field across the scan area. 36× 36 cm² scan area was chosen in the middle of the prototype. The loop was fixed 10 mm above the prototype surface. The scanning step was 15 mm. The magnetic fields in the absence of the prototype were measured as a reference. Then, using transformation of the measured near fields to the far-field zone and post-processing, we obtained the transmittances[3].

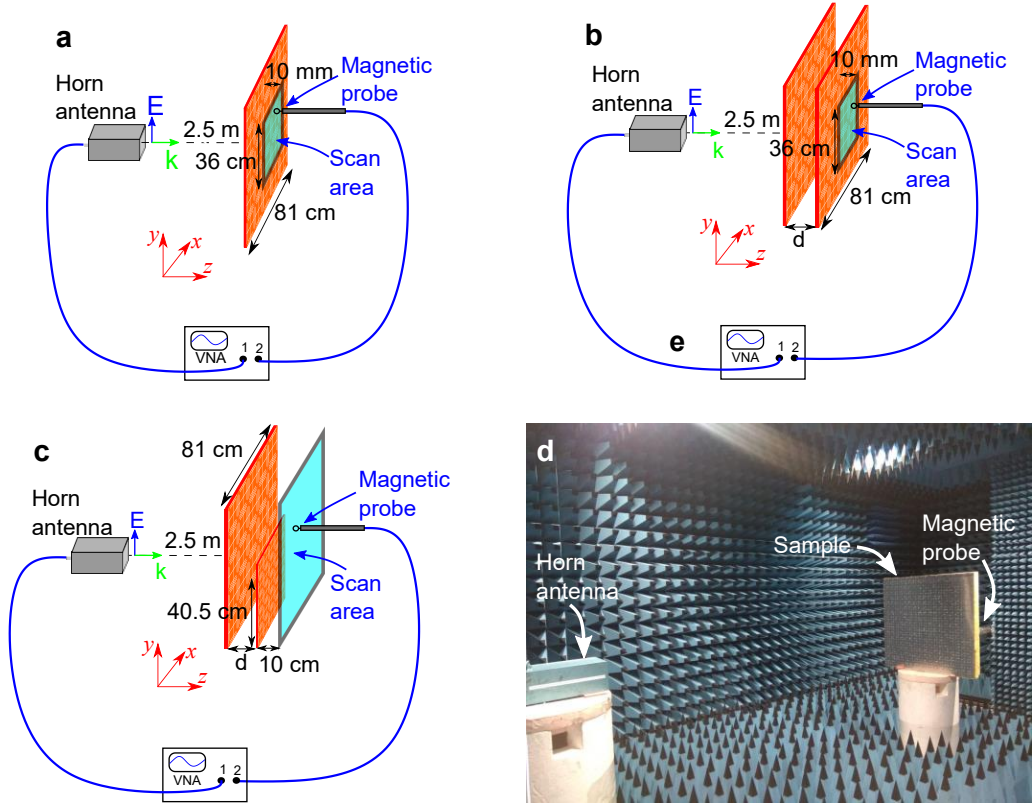

**Fig. S4| Experimental setups.** Schematic views of the experimental setups for measurements of (a) single layer ENMZ metasurface transmittance with the size of $81 \times 81$ cm ($18 \times 18$ unit cells), (b) EMNZ-air-EMNZ structure transmittance with the size of $81 \times 81$ cm ($18 \times 18$ unit cells), and (c) scanning of near magnetic field distributions for EMNZ-air-small EMNZ structure. In panel (c) the EMNZ metasurface size is $81 \times 81$ cm ($18 \times 18$ unit cells) and the EMNZ slab size is $40.5 \times 40.5$ cm ($9 \times 9$ unit cells). (**d**) Photo of the experimental setup for near magnetic field scanning.

We also measured the magnetic near-field distribution for *2L* and *3L* structures for different positions of the EMNZ slab. A schematic view of the near-field mapping setup is illustrated in Fig. S4c. The *x* component of the magnetic field ($H_x$) was scanned over the area of $80 \times 80$ cm² with 4 cm step. The loop was fixed 10 cm above the EMNZ slab surface. Fig. S4d shows a photo of the experiment setup.

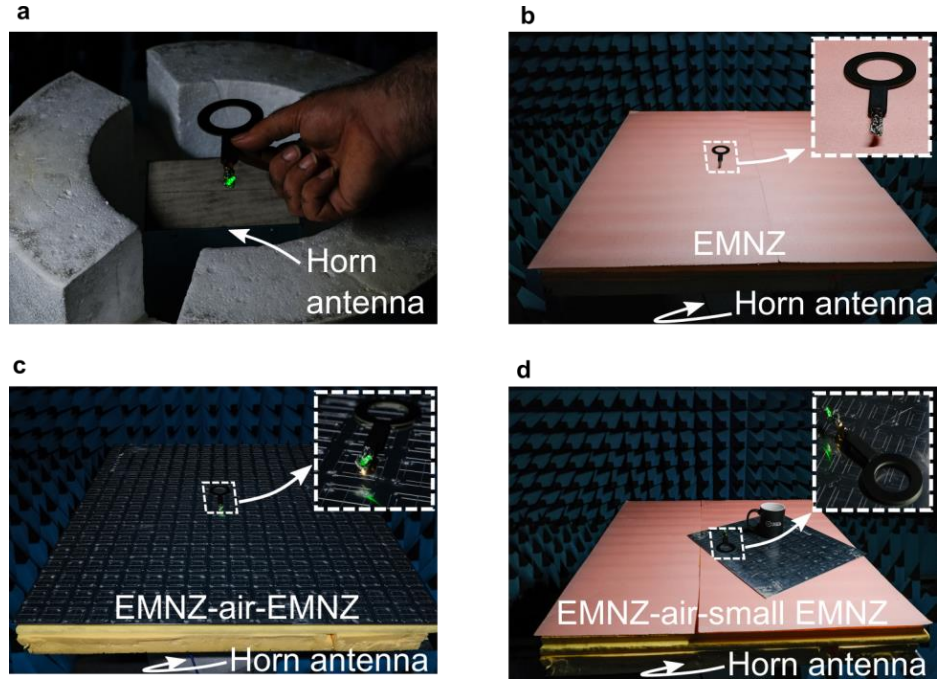

**Fig. S5| LED demonstration. a,** power transmission from a horn antenna to a loop receiver loaded by a LED in free space. **b,** adding 2L structure between the transmitter antenna and receiver switches off the LED. **c,** demonstration of power transfer from the horn antenna to the LED loaded receiver for 3L structure. **d,** demonstration of targeted power transmission using *3L* structure for rotated EMNZ slab.

At the final step, a set of LED demonstration experiments were performed in order to visualize the targeted power transfer concept. The sinusoidal signal at a fixed frequency was produced by a Rohde & Schwarz SMB100A signal generator. Then it was amplified with an Agilent 83020A power amplifier up to 1W and stimulated the horn antenna. A 9 kHz-6 GHz Foresightloop probe was used as a receiver. A simple full-bridge rectifying network composed of B240A-E3 Schottky diodes connected to the output of the receiver was used to convert the AC signal to DC one. Next, the network was loaded by a LED to visualize the power transfer. If the receiver is placed above the horn antenna in free space (see Fig. S5a), the LED receives the needed power and is shining. For the 2L structure (with the size of 81×81 cm$^2$) located 5cm above the horn antenna, we observe that the receiver located in the central position 8 cm above the 2L structure does not provide any power. As expected, the LED is off (Fig. S5b). It means no electromagnetic wave can tunnel the 2L structure. As soon as we added the EMNZ slab (3L structure), the power tunnelling occurred, and the LED was shining (Fig. S5c). We also proved the same scenario for the 3L structure for EMNZ slab 40.5×40.5 cm$^2$) and with an angle misalignment made by rotating the EMNZ slab (Fig. S5d). Again, the LED is shining, and the power tunnelling effect is still there.